\documentclass[aps,prb,reprint,superscriptaddress,showpacs]{revtex4-1}
\usepackage{graphicx}
\usepackage{textcomp}
\usepackage{epstopdf}
\usepackage{gensymb}

\begin{document}

\title{Tetragonal magnetic phase in Ba$_{1-x}$K$_x$Fe$_2$As$_2$ from x-ray and neutron diffraction }

\author{J. M. Allred}
\email[]{jallred@anl.gov}
\affiliation{Materials Science Division, Argonne National Laboratory, Argonne, IL 60439-4845, USA}
\author{S. Avci}
\affiliation{Materials Science Division, Argonne National Laboratory, Argonne, IL 60439-4845, USA}
\affiliation{Department of Materials Science and Engineering, Afyon Kocatepe University, 03200 Afyon, Turkey}
\author{D. Y. Chung}
\author{H. Claus}
\affiliation{Materials Science Division, Argonne National Laboratory, Argonne, IL 60439-4845, USA}
\author{D. D. Khalyavin}
\author{P. Manuel}
\affiliation{ISIS Pulsed Neutron and Muon Source, Rutherford Appleton Laboratory, Chilton, Didcot OX11 0QX, United Kingdom}
\author{K. M. Taddei}	
\affiliation{Materials Science Division, Argonne National Laboratory, Argonne, IL 60439-4845, USA}
\affiliation{Physics Department, Northern Illinois University, DeKalb, IL 60115, USA}
\author{M. G. Kanatzidis}
\affiliation{Materials Science Division, Argonne National Laboratory, Argonne, IL 60439-4845, USA}
\affiliation{Department of Chemistry, Northwestern University, Evanston, IL 60208-3113, USA}
\author{S. Rosenkranz}
\author{R. Osborn}
\affiliation{Materials Science Division, Argonne National Laboratory, Argonne, IL 60439-4845, USA}
\author{O. Chmaissem}
\affiliation{Materials Science Division, Argonne National Laboratory, Argonne, IL 60439-4845, USA}
\affiliation{Physics Department, Northern Illinois University, DeKalb, IL 60115, USA}

\date{\today}

\begin{abstract}
Combined neutron and x-ray diffraction experiments demonstrate the formation of a low-temperature minority tetragonal phase in Ba$_{0.76}$K$_{0.24}$Fe$_2$As$_2$ in addition to the majority magnetic, orthorhombic phase. A coincident enhancement in the magnetic ($\frac{1}{2}$ $\frac{1}{2}$ 1) peaks shows that this minority phase is of the same type that was observed in Ba$_{1-x}$Na$_x$Fe$_2$As$_2$ ($0.24 \leq x \leq 0.28$), in which the magnetic moments reorient along the $c$-axis. This is evidence that the tetragonal magnetic phase is a universal feature of the hole-doped iron-based superconductors.
\end{abstract}


\maketitle

\section{Introduction}

A key to understanding the pairing mechanism in the superconducting state of the iron pnictide and chalcogenide superconductors is characterizing the nature of the electronic interactions that are responsible for the competing spin-density-wave (SDW) state. The origin of the magnetic order, which is coupled to an orthorhombic distortion of the high-temperature tetragonal lattice that is often labelled `nematic' order, is still debated, with a range of theoretical treatments that range from localized orbital models to weak-coupling itinerant models based on Fermi surface nesting.\cite{Kontani_orbital_2012,Fernandes:2014jf,Dai:2012em}

Since it seems that, until now, each class of models could be modified to comport with the measured properties, solely studying the principal SDW state is apparently not enough to settle the dispute. Recently we have shown that a second tetragonal magnetic phase appears near the end of the SDW dome in hole-doped Ba$_{1-x}$Na$_x$Fe$_2$As$_2$. This was called the ``$C_4$'' phase to distinguish it from the more usual stripe SDW phase, which has ``$C_2$'' symmetry.\cite{Avci:2014fp, Avci:2013iua,Avci:2011cj} The magnetic structure of the $C_4$ phase was later shown to involve a reorientation of the spins from in-plane to out-of-plane.\cite{Wasser:2015fw,Khalyavin:2014dg}  The reorientation temperature, $ T_{\rm{r}}$, occurs well above $T_{\rm{c}}$, indicating that it does not arise from a coupling of the superconducting order parameter with the other order parameters as was observed in Ba(Fe$_{1-x}$Co$_x$)$_2$As$_2$,\cite{Nandi:2010ea} but instead must be a manifestation of changes in the coupling between iron atoms.  The fact that the system returns to the tetragonal symmetry of the paramagnetic phase, while remaining magnetic, puts new constraints on the set of plausible electronic ordering mechanisms.  In particular, a group theory analysis shows that it may be possible to distinguish between itinerant and quasi-local orbital models through the observation of orbital order in the $C_4$ phase.\cite{Khalyavin:2014dg}

Characterizing the $C_4$ SDW phase is therefore a primary concern in settling the debate on the origin of the parent $C_2$ SDW phase. It is important to establish whether this phase is unique to the Ba$_{1-x}$Na$_x$Fe$_2$As$_2$ system, or whether it is observed in other related systems, in order to decide if the inferred physics of these specific compounds are atypical or more general. Related families such as the hole-doped Ba$_{1-x}$K$_x$Fe$_2$As$_2$ have been extensively studied in earlier publications.\cite{rotter_superconductivity_2008-1, Avci:2012ha}   Here we return our attention to the region near the edge of the dome of  Ba$_{1-x}$K$_x$Fe$_2$As$_2$ using fine temperature control with both high-resolution x-rays and high-intensity neutrons.  The combined analysis shows that the reentrant $C_4$ tetragonal phase is indeed present in the Ba$_{1-x}$K$_x$Fe$_2$As$_2$ phase diagram at $x$ = 0.24, but not below $x$ = 0.22, indicating that it is a common feature of the hole-doped `122' iron compounds. It is only present as a minority phase below a first-order transition, consistent with the delicate energy balance between the $C_2$ and $C_4$ phases predicted by itinerant spin-nematic theory,\cite{Fernandes:2014jf} and is rapidly suppressed below the superconducting transition. We had earlier reported that the $C_4$ phase in Ba$_{1-x}$Na$_x$Fe$_2$As$_2$ exhibited a stronger competition with superconductivity than the $C_2$ phase,\cite{Avci:2014fp} but this is the first time we have observed the complete suppression of the $C_4$ phase below $T_{\rm{c}}$.

\section{Techniques}
The polycrystalline samples reported in Ref \rm{\citenum{Avci:2012ha}} were used in the new measurements.   Powders were prepared by combining stoichiometric amounts of BaAs, KAs, and Fe$_2$As in sealed Nb tubes, which were in turn sealed in quartz tubes and fired at $1050~\celsius$.  More details can be found in the original publication.\cite{Avci:2012ha} 
  Powder x-ray diffraction experiments (PXRD) were measured at the Advanced Photon Source, Argonne National Laboratory, on beamline 11-BM-B using the liquid helium cryostat ($\lambda =0.413429$ \r{A}). The sample was prepared for measurement by dusting the outside of a greased kapton capillary with sample powder. Rietveld refinements were performed using General Structure Analysis System (GSAS)\cite{gsas} and  the graphical user interface, EXPGUI.\cite{expgui} Powder neutron diffraction (PND) experiments were performed at ISIS, Rutherford Appleton Laboratory, on the Wish beamline.

\begin{figure}
 \includegraphics[width = \columnwidth]{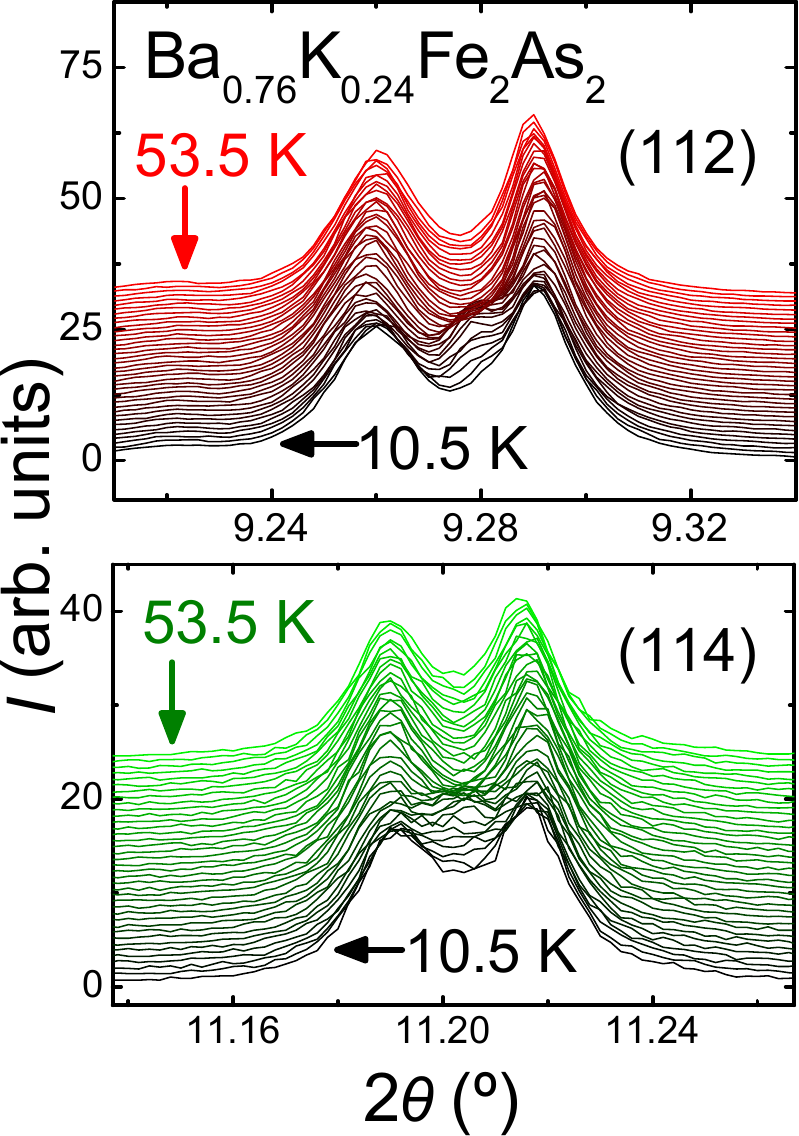}
 \caption{Overlaid x-ray diffraction scans of Ba$_{0.76}$K$_{0.24}$Fe$_2$As$_2$ at approximately uniform increments from 10.5 to 53.5 K. The top panel shows the (112) reflection and the bottom shows the (114) reflection. The split peak in the orthorhombic phase corresponds to the ($11l$) and ($\bar{1}1l$) components ($I$-cell), which are the ($20l$) and ($02l$) peaks in the conventional F-centered cell.  \label{xhist1}}
 \end{figure}

 \begin{figure}
 \includegraphics[width = \columnwidth]{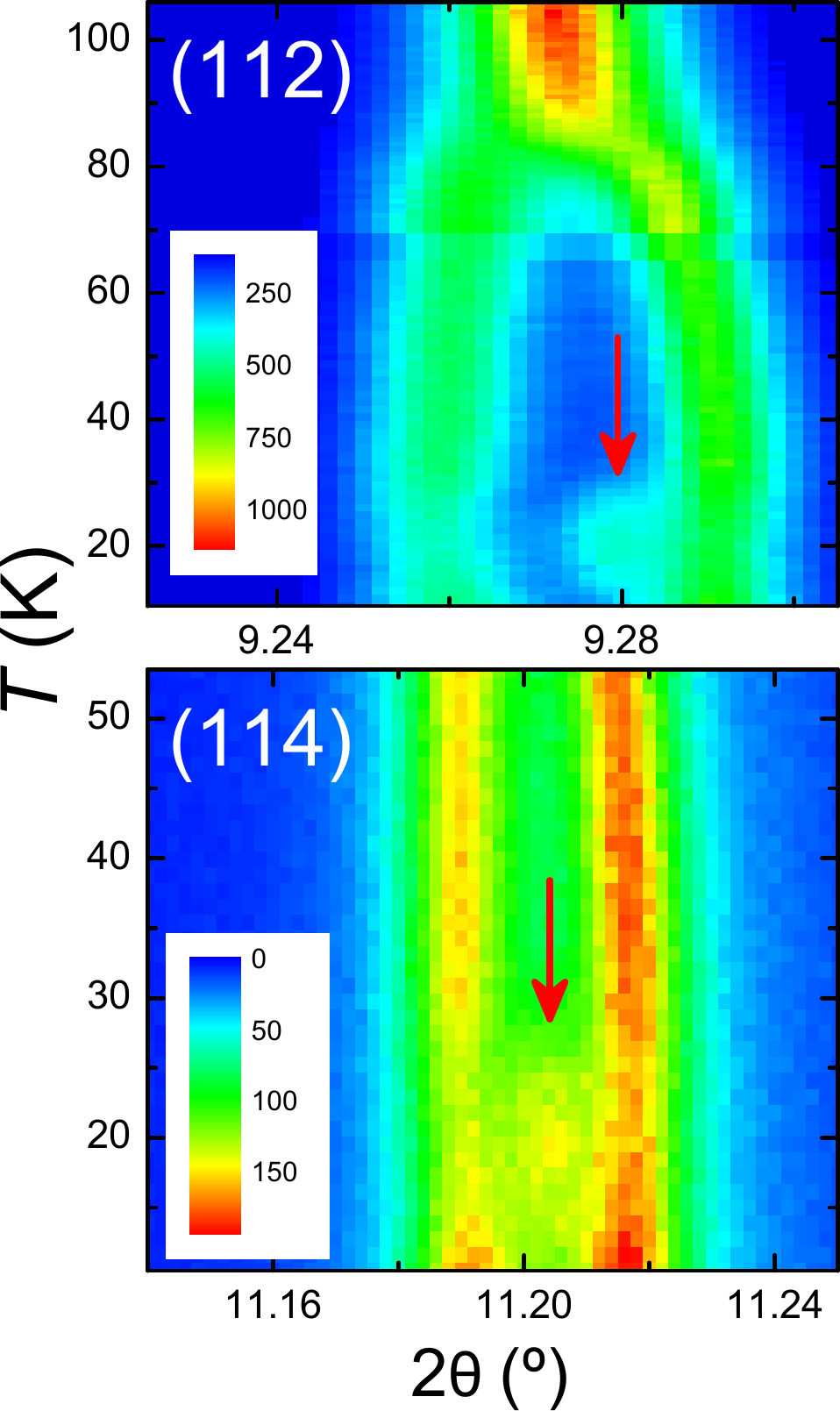}
 \caption{False color map drawn using the x-ray diffraction data on Ba$_{0.76}$K$_{0.24}$Fe$_2$As$_2$. The top panel shows the temperature range from 10 to 120 K of the (112) reflection, which splits into the (202) and (022) orthorhombic peaks. The bottom panel shows the (114) (corresponding to the (204) and (024) components of the orthorhombic cell) from 10 to 53 K. The arrows point to the peak corresponding to the minority tetragonal phase.\label{xhist2}}
 \end{figure}

 \begin{figure}
 \includegraphics[width = \columnwidth]{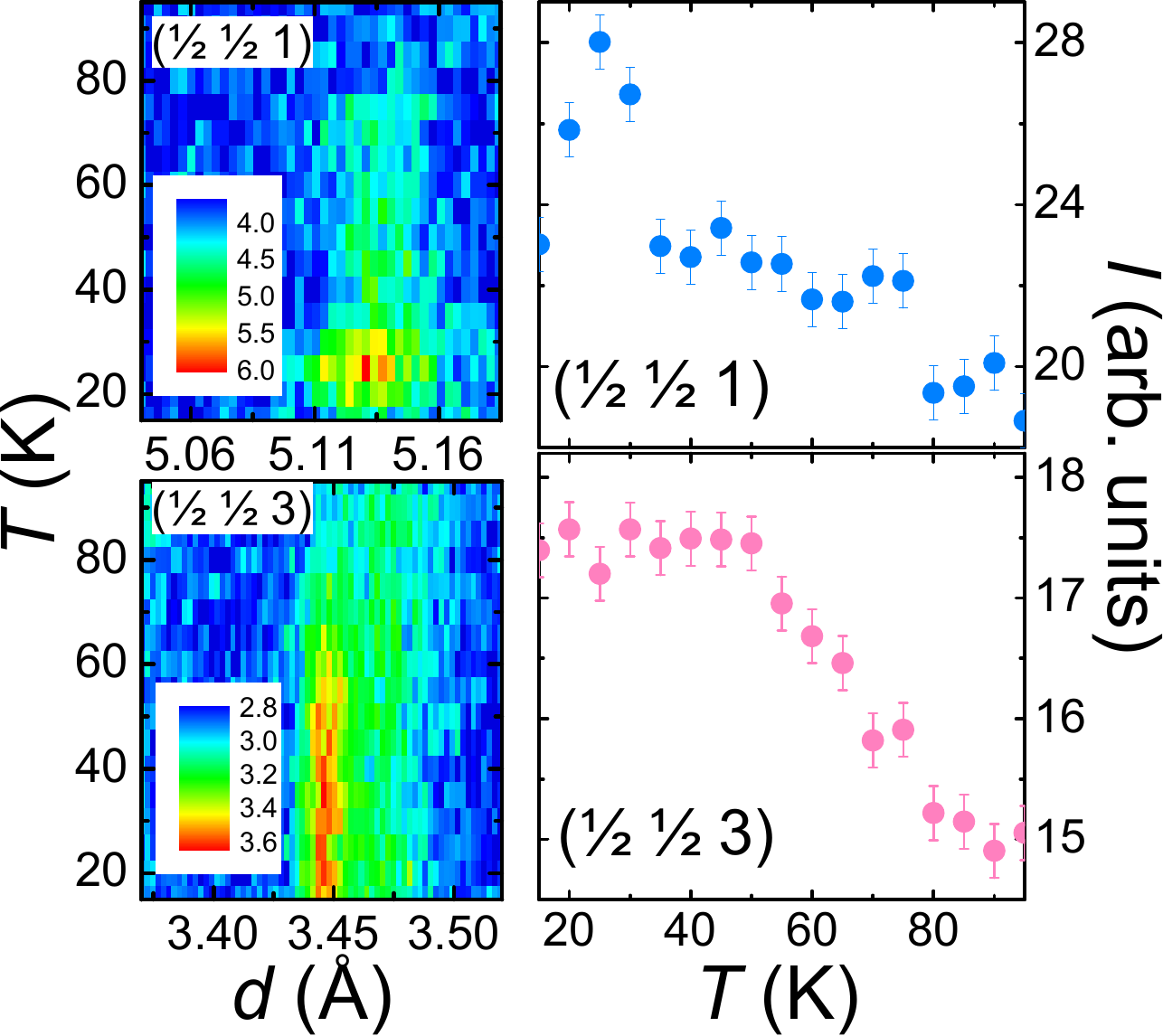}
 \caption{Left panels: Temperature-dependent color maps depicting the magnetic peak intensity of the ($\frac{1}{2}$ $\frac{1}{2}$ 1) (top) and ($\frac{1}{2}$ $\frac{1}{2}$ 3) (bottom) reflections. The right panels are peak intensities integrated from a 5-point-wide cut through the peak of interest.\label{nehist}}
 \end{figure}

\section{Results and discussion}
The diffraction data are summarized in Figures \ref{xhist1}, \ref{xhist2} and \ref{nehist}. The peak indices are given for the body-centered tetragonal cell in all cases. The cascaded diffractograms of the x-ray diffraction data (Figure \ref{xhist1}) shows the splitting of the (112) and (114) reflections, with clear evidence of a minority phase growing in and disappearing again on warming from 10 ($T_{\rm{r,1}}$ to 30 K ($T_{\rm{r,2}}$) between these two reflections.  The same data can also be visualized using a false color map, as depicted in Figure \ref{xhist2}.  The neutron diffraction data (Figure \ref{nehist}) shows an evolution in the magnetic intensity within the same region for the ($\frac{1}{2}\frac{1}{2}1$) reflection. Little change is observed in the ($\frac{1}{2}\frac{1}{2}3$) peak.  Our previous study on Ba$_{1-x}$Na$_x$Fe$_2$As$_2$ showed that the $C_4$ phase exhibits a 10-fold increase in intensity of the ($\frac{1}{2}\frac{1}{2}1$) magnetic peak, and a slight reduction in the ($\frac{1}{2}\frac{1}{2}3$) magnetic peak, resulting from the spin reorientation.  Here the magnitude change in the ($\frac{1}{2}\frac{1}{2}1$) is consistent with an approximately 10\% $C_4$ phase fraction.

A two-phase Rietveld refinement was used to model the minority phase in the 11BM data.   For two-overlapping phases of such similar structure, it is difficult to refine the lattice parameters for the minority phase reliably, and similar R values are obtained regardless of whether certain parameters of the tetragonal phase are constrained or allowed to refine freely. For example, freely refining the lattice parameters of the minority phase gives obviously incorrect peak positions, due to convolution with small features in the majority phase peak's shoulders that are not perfectly modeled. These values also covary with the relative phase fraction, making a unique solution unattainable. As detailed below, several assumptions and approximations must be made in order to minimize systematic errors and to produce a model that is physically meaningful.

To start, the $c$-axes of both phases ($c_1$ and $c_2$, for the majority orthorhombic and minority tetragonal phases, respectively) were constrained to be equivalent, since no extra broadening of the ($00l$) reflections was observed when comparing the two-phase and one-phase temperature regions. The shoulders in the peaks split by the orthorhombic distortion (such as what is shown in Figure \ref{xhist1}) can then be used to define the other cell axis of the minority phase, $a_2$, which allows the free refinements of $a_1$ and $b_1$, giving good agreement with the data.  Thermal and peak profile parameters of both phases are constrained to be nearly equivalent, which allowed the weight fractions to be refined self-consistently.   Using this method, the final two-phase refinement models have all of the crystallographic parameters of the orthorhombic phase freely refined, while, for the tetragonal phase, only the scale factor is refined by GSAS. 

\begin{figure}
 \includegraphics[width = \columnwidth]{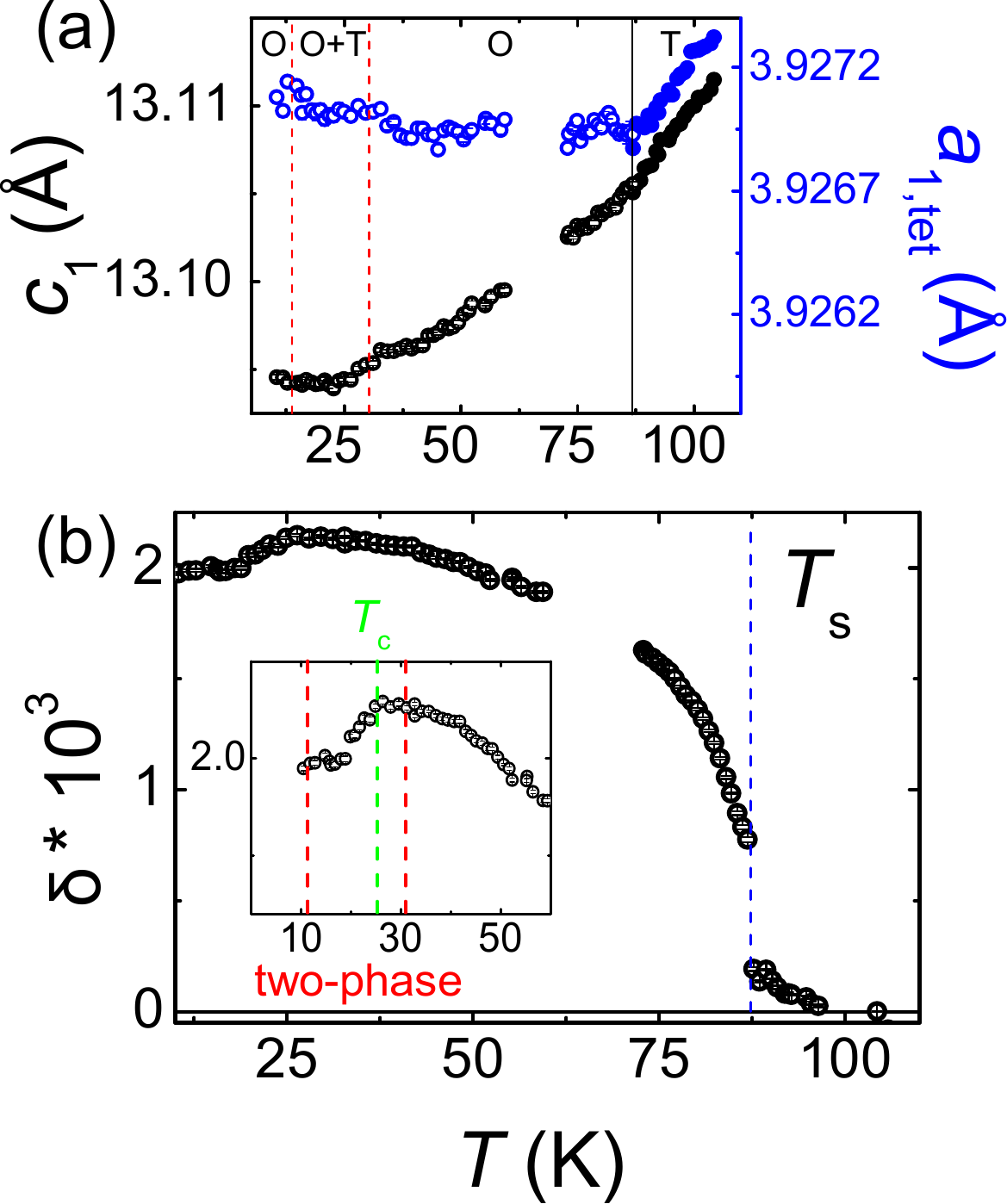}
 \caption{(a) Refined lattice parameters for the majority phase in  Ba$_{0.76}$K$_{0.24}$Fe$_2$As$_2$ determined from x-ray powder diffraction. Black circles represent $c_1$, and blue circles represent $a_{\rm{1,tet}}$ (see text). Orthorhombic and tetragonal fits were used below and above 87 K, respectively. (b) The orthorhombic order parameter ($\delta_1= \frac{a_1-b_1}{a_1+b_1}$) is plotted from refinements where the majority phase is refined as orthorhombic at all temperatures for comparison. Inset is a detailed view around $T_{\rm{c}}$ and $T_{\rm{r}}$ \label{refinements1}}
 \end{figure}

\begin{figure}
 \includegraphics[width = \columnwidth]{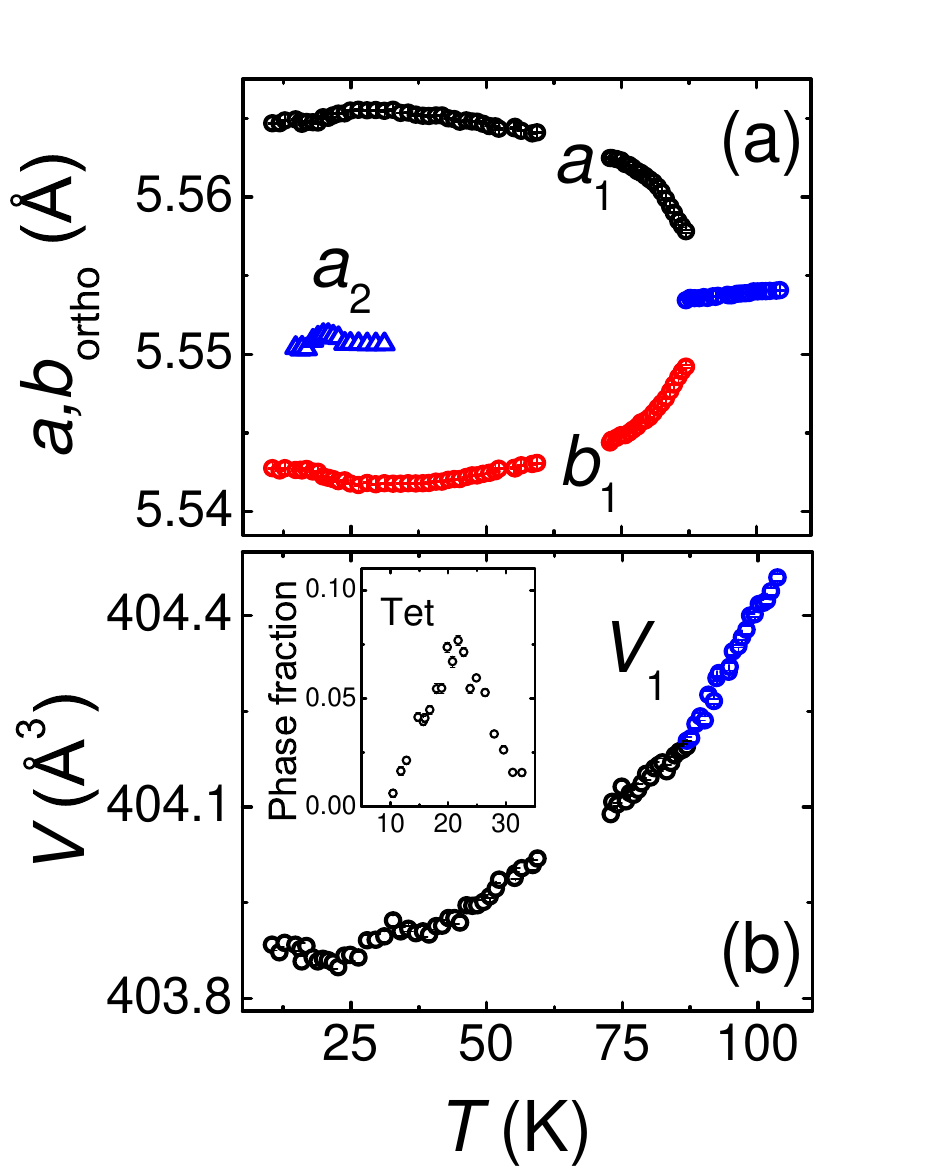}
 \caption{(a) The splitting of the $a$ (black circles) and $b$ (red circles) lattice parameters in the orthorhombic phase, along with the minority phase tetragonal $a_2$ (blue triangles) lattice parameter.  Tetragonal lattice parameters are scaled by $\sqrt{2}$. (b) The volume of the majority phase ($V_1$), with blue and black circles being used to demarcate the transition from tetragonal to orthorhombic models, respectively The inset is the refined phase fraction of the minority phase, when present.  \label{refinements2}}
 \end{figure}

The temperature dependence of the majority phase is summarized in Figures \ref{refinements1} and \ref{refinements2}.  The refined lattice parameters of the majority (orthorhombic) phase do not appear to be affected by the presence of the minority (tetragonal) phase. For example, the orthorhombic order parameter shows the usual discontinuity at $T_c$ (26 K), but there is no evidence of a change in slope corresponding to the (dis)appearance of the tetragonal phase. This is further evidence that the transition from orthorhombic $C_2$ phase to the magnetic $C_4$ phase is of first order and that the two phases are microscopically decoupled. The structural transition at $T_{\rm{s}}$ is also clearly in the primitive basal plane lattice parameter (here, called $a_{\rm{tet}}$) and orthorhombic order parameter, $\delta = \frac{a_1-b_1}{a_1+b_1}$ (Figure \ref{refinements1}a and b, respectively). The in-plane lattice parameter, $ a_{\rm{tet}}$, is calculated from the orthorhombic phase by transforming the conventional $F$-centered cell back to the $I$-centered one, \textit{via} $a_{1,\rm{tet}} = \sqrt{(a_1^2 + b_1^2)/2}$.  Below $T_{\rm{N}}$, this value is clearly enhanced, as is typical of hole-doped 122 iron-pnictides,\cite{PhysRevB.90.104513} while the c-axis shows little change. The volume anomaly at $T_{\rm{N}}$ is primarily seen as a subtle change in slope.

The properties of the minority phase are also plotted in Figure \ref{refinements2}. What is observed is a small tetragonal phase fraction that reaches a maximum ($\sim 8\%$) between 20.8 and 22.7 K, and becomes indistinguishable from background below 12 K and above 30 K. The in-plane lattice parameter appears to be rather smaller than the primary phase (Figure \ref{refinements2}).  The phase fraction derived from the x-ray data peaks at the same temperature as the enhancement in the magnetic peak intensity of the ($\frac{1}{2}$ $\frac{1}{2}$ 1) reflection in neutron data. This agrees well with the interpretation that this minority phase is the same tetragonal magnetic phase that was observed in Ba$_{1-x}$Na$_x$Fe$_2$As$_2$.\cite{Avci:2014fp}

 \begin{figure}
 \includegraphics[width = \columnwidth]{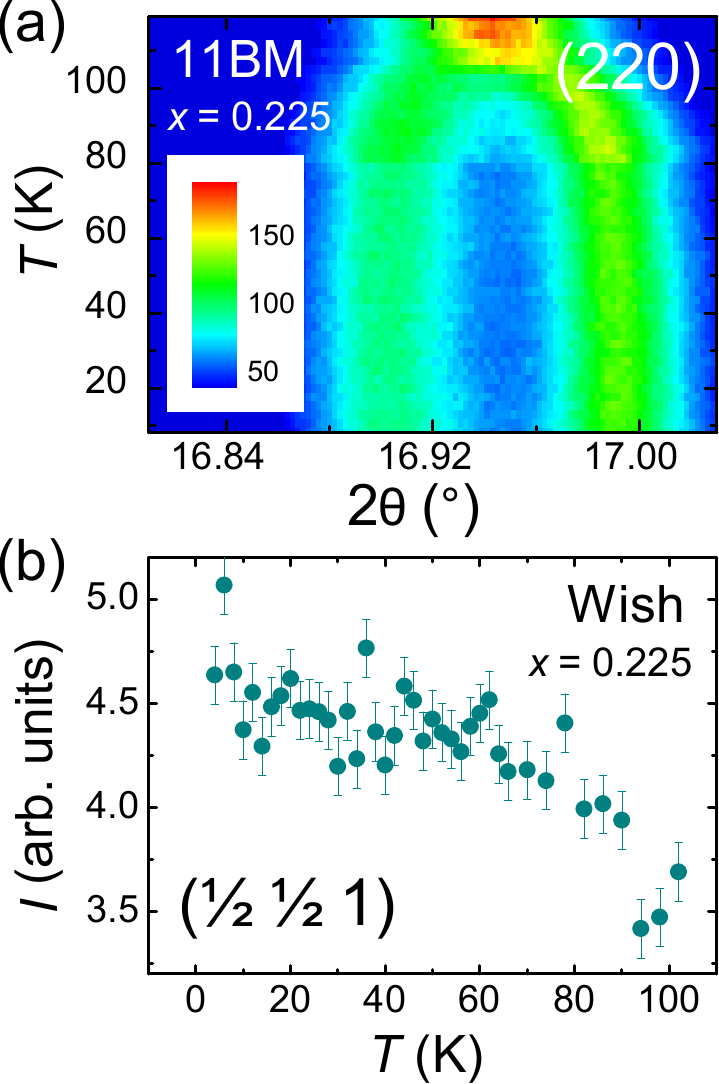}
 \caption{Powder diffraction data from Ba$_{0.775}$K$_{0.225}$Fe$_2$As$_2$. (a) X-ray (11BM) diffractogram of the (220) reflection. (b) Integrated intensity across the ($\frac{1}{2}\frac{1}{2}1$) reflection from the neutron (Wish) diffraction data. \label{otherdata}}
 \end{figure}

 \begin{figure}
 \includegraphics[width = \columnwidth]{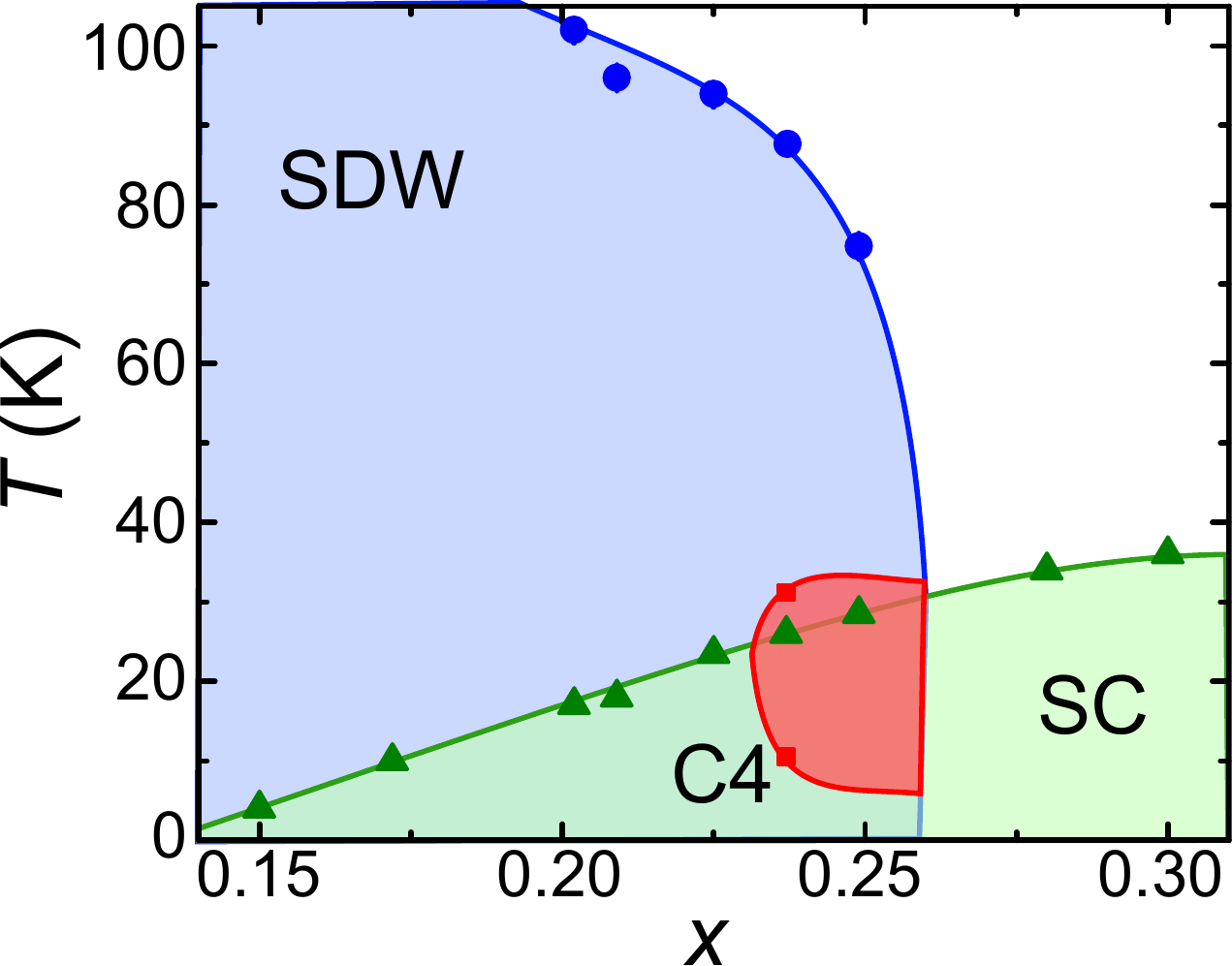}
 \caption{Phase diagram of Ba$_{1-x}$K$_x$Fe$_2$As$_2$ revised from Ref. \citenum{Avci:2012ha} to include the new diffraction data from this study. \label{phasediagram}}
 \end{figure}

In order to draw a modified phase diagram for the Ba$_{1-x}$K$_x$Fe$_2$As$_2$ family, some understanding of the limits to the $C_4$ region need to be established. Similar high-resolution x-ray and high-intensity neutron diffraction experiments on a $x$ = 0.225(10) sample, with a temperature spacing of 2 K, show no evidence of a low-T tetragonal phase nor of any spin reorientation (Figure \ref{otherdata}a and b, respectively), which sets a lower limit. While our previous report of high-resolution neutron diffraction \cite{Avci:2012ha} provides insufficient data to rule out the possibility that a small tetragonal phase appears below 40 K at $x=0.25$,  recent uniaxial dilatometry measurements provide evidence of a return to tetragonal symmetry in a narrow temperature range all the way up to the edge of the magnetic dome (in compositional space)\cite{2014arXiv1412.7038B}. While the application of stress may change the properties enough to shift phase boundary lines (for example, see Hassinger\emph{ et al.}\cite{Hassinger:2012ki}), this is consistent with our results, so the phase diagram of Ba$_{1-x}$K$_x$Fe$_2$As$_2$ is drawn to be consistent with this scenario in Figure \ref{phasediagram}. The complete suppression of the $C_4$ phase below $T_{\rm{c}}$, shown in the inset to Fig. \ref{refinements2}b, is consistent with our earlier observation that it competes more strongly with superconductivity than the $C_2$ phase.\cite{Avci:2014fp}

\section{Conclusion}
The combined neutron and x-ray powder diffraction on Ba$_{0.76}$K$_{0.24}$Fe$_2$As$_2$ shows evidence of a second, minority phase below 30 K. The structural and magnetic features of the secondary phase indicate that it is the same as the previously reported tetragonal magnetic phase in Ba$_{1-x}$Na$_x$Fe$_2$As$_2$, \cite{Avci:2014fp} but with a reduced stability range, possibly due to the larger cation size. This provides evidence that the $C_4$ phase is a universal feature of the hole-doped 122 family of iron-based superconductors. The resulting phase diagram is similar to the one recently reported by B\"ohmer \textit{et al}\cite{2014arXiv1412.7038B}, although they claim that the tetragonal phase fraction is 100\%. The fact that we observe it as a minority phase suggests that the stability of the $C_4$ phase is extremely sensitive to subtle changes in sample composition and measurement conditions. We have proposed that the $C_4$ phase is evidence for itinerant spin-nematic theory, in which the coupled magnetic and structural transitions are due to magnetic fluctuations caused by Fermi surface nesting. This theory predicts that the $C_2$ and $C_4$ phases have very similar free energies close to the suppression of $C_2$ order,\cite{Avci:2014fp} which is consistent with the delicate stability of the $C_4$ phase. In this model, the $C_4$ SDW results from a simultaneous coupling between Fermi surfaces along two in-plane directions, rather than just one in the $C_2$ phase, providing a natural explanation for the stronger phase competition with superconductivity evident in the phase diagram.

\begin{acknowledgments}
This work was supported by the U.S. Department of Energy, Office of Science, Materials Sciences and Engineering. This research used resources of the Advanced Photon Source, a U.S. Department of Energy (DOE) Office of Science User Facility operated for the DOE Office of Science by Argonne National Laboratory under Contract No. DE-AC02-06CH11357, and was aided by the 11-BM beam scientist M. Suchomel. Experiments at the ISIS Pulsed Neutron and Muon Source were supported by a beam time allocation from the Science and Technology Facilities Council.
\end{acknowledgments}

\bibliography{k122lib}

\end{document}